\documentclass[aps,prl,twocolumn,superscriptaddress]{revtex4}

\usepackage{graphicx}
\begin{document}

%

\preprint{Ver 3.0 8.11.2001}

\title{
Field dependent anisotropy change in a supramolecular Mn(II)-[3$\times$3] grid
}

\author{
O. Waldmann
}
\email[Corresponding author.\\E-mail: ]{waldmann@physik.uni-erlangen.de}
\affiliation{
Physikalisches Institut III, Universit\"at Erlangen-N\"urnberg, D-91058 Erlangen, Germany
}

\author{
L. Zhao
}
\author{
L. K. Thompson
}
\affiliation{
Department of Chemistry, Memorial University, St. John's, NF, Canada A1B 3X7
}

\date{\today}

\begin{abstract}
The magnetic anisotropy of a novel Mn(II)-[3$\times$3] grid complex was investigated by means of high-field torque magnetometry. Torque vs. field curves at low temperatures demonstrate a ground state with $S > 0$ and exhibit a torque step due to a field induced level-crossing at $B^* \approx 7.5\,$T, accompanied by an abrupt change of magnetic anisotropy from easy-axis to hard-axis type. These observations are discussed in terms of a spin Hamiltonian formalism.
\end{abstract}


\maketitle

%

The design of clusters of magnetic metal centers with novel magnetic properties has become a major goal in the area of nanoscale materials. A particularly interesting class of compounds is that of the supramolecular [$N \times N$] grid structures \cite{grids}. The essentially flat, square-matrix-like arrangement of exactly $N^2$ metal ions realized in these complexes is not only aesthetically pleasant but also suggests applications as, e.g., molecular storage devices \cite{grids}. From the magnetic perspective, the grids may be regarded as molecular model systems for magnets with extended interactions on a square lattice which are, e.g., relevant in the context of high-temperature superconductors.

Numerous [2$\times$2] grids containing magnetic metal ions have been created, and magnetic studies have shown a remarkable variety of their magnetic properties \cite{Fe22g,Cu22gLT,Co22g,Ni22g}. For a [2$\times$2] square of spins the distinction between a "two-dimensional" grid and a "one-dimensional" ring is really semantic, but for a [3$\times$3] grid it is truly "two-dimensional". However, the first [3$\times$3] grid complexes with magnetic ions could be synthesized only very recently \cite{Cu33gLT,Mn33gLT}.

In this work, we investigated the Mn-[3$\times$3] grid cluster [Mn$_9$(2POAP-2H)$_2$]$^{6+}$ (Fig. 1) by means of magnetization and high-field torque measurements, focusing on its anisotropic properties. At low temperatures an unusual field dependence of the magnetic anisotropy was observed: With increasing field, the torque signal exhibits a clear step where the magnetic anisotropy changes its sign abruptly from easy-axis to hard-axis type.

\begin{figure}
\includegraphics{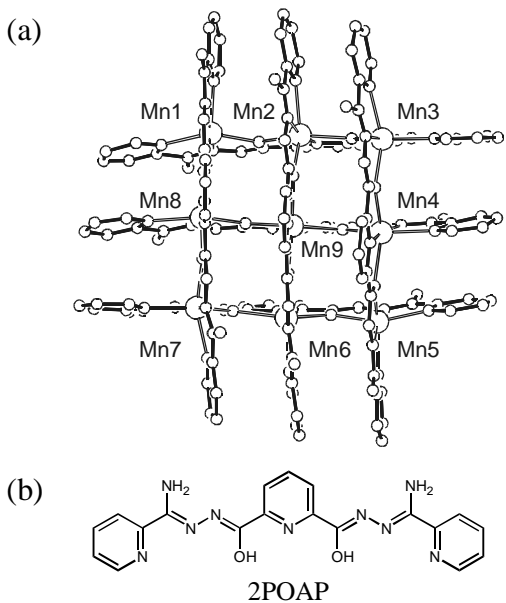}
\caption{
(a) Structural representation of the cation [Mn$_9$(2POAP-2H)$_6$]$^{6+}$ and (b) sketch of the ligand.
}
\end{figure}

%

Single crystals of [Mn$_9$(2POAP-2H)$_6$] (ClO$_4$)$_6$ $\cdot$ 3.57 MeCN $\cdot$ H$_2$O were synthesized as reported \cite{Mn33gLT}. They crystallize in the space group $C_2/c$. The cation [Mn$_9$(2POAP-2H)$_2$]$^{6+}$ exhibits a slightly distorted $S_4$ molecular symmetry with the $C_2$ symmetry axis being perpendicular to the plane of the cluster. The distances of the Mn(II) ions are in between $3.88 - 4.02 \, \AA$ with an average of $3.93 \, \AA$. The crystallographic unit cell contains eight positions. They can be divided into two sets since each group of four are magnetically equivalent by symmetry. The molecular $C_2$ symmetry axes are parallel for all positions, the two sets of magnetically non-equivalent positions are twisted by 31$^\circ$ around the $C_2$ axes.

The magnetic moment of powder samples, prepared by air drying of crystals, was measured with a SQUID magnetometer (Quantum Design) as described in Ref. \onlinecite{Ni22g}. The weights of the samples were ca. 4$\,$mg. Data were corrected for ligand diamagnetism and TIP. Susceptibility was determined from measurements at a field of 0.1$\,$T. The torque of single crystals was measured with a homemade cantilever device micromachined from crystalline silicon (inset of Fig. 3), inserted into a 15$\,$T/17$\,$T $^4$He cryomagnet. For details see Ref. \onlinecite{CsFe8}. Resolution was 10$^{-11} \,$Nm, non-linearity was less than 1\%, accuracy of in-situ alignment was $1^\circ$. The weights of the two crystals investigated were ca. 25$\, \mu$g. Crystals were
mounted on the cantilever with the $C_2$ symmetry axis parallel to the $z$-axis. Calibration of the signal is accurate to $\pm 25 \%$.

%

The planar structure of the Mn-[3$\times$3] grid suggests an almost uniaxial magnetic behavior, with the uniaxial axis coinciding with the molecular $C_2$ symmetry axis. This agrees with findings for other planar molecules \cite{Co22g,Ni22g,CsFe8,Cu33g,XFe6}, but was also checked experimentally. The in-plane anisotropy of the grid was found to be at best 15$\%$ of the main $||$-$\perp$ anisotropy, i.e. can be neglected. This estimate includes the virtual reduction of an in-plane anisotropy due to the 31$^\circ$ twist of the two sets of clusters in the crystal structure.

Figure 2 presents the temperature dependence of the magnetic susceptibility and the magnetization
curve at 1.8$\,$K for powder samples. The decrease of $\chi T$ with decreasing temperature indicates the presence of antiferromagnetic interactions. Zero-field-cooled and field-cooled curves (measured at 2$\,$mT) were identical. Thus, intermolecular interactions are negligibly small as expected from the minimum distance between grids of $8 \, \AA$ in the crystal structure. The room temperature value of $\chi T$ of 63.3$\, N_{A} \, \mu_B \,$T$^{-1} \,$K is consistent with nine high-spin Mn(II) centers. The $\chi T$ values at low temperatures are close to 7.84$\, N_{A} \, \mu_B \,$T$^{-1} \,$K pointing towards a $S = 5/2$ ground state. At the lowest temperatures, $\chi T$ drops further signaling a weak magnetic anisotropy of the ground state. The magnetization curve [Fig. 2(b)] confirms $S = 5/2$ for the ground state. The increase of magnetic moment at higher fields indicates the presence of low-lying excited states.

\begin{figure}
\includegraphics{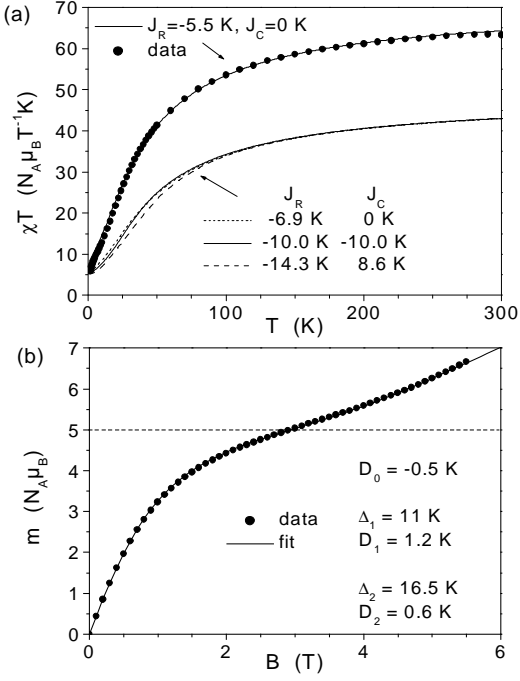}
\caption{
(a) Temperature dependence of the magnetic susceptibility, drawn as $\chi T$, of Mn-[3$\times$3] 
powder (full circles). The solid line represents a fit with $J_R=-5.5\,$K, $J_C=0$, $g=2$. The 
three other curves correspond to grids with spin-2 centers and coupling constants as given. (b) Field dependence of the magnetic moment at $T=1.8\,$K of Mn-[3$\times$3] powder. The solid line 
represents a fit with the model described in the text with parameters given in the panel.
}
\end{figure}

A typical result for the field dependence of the torque at $T = 1.75 \,$K is shown in Fig. 3(a). The increase of signal at low fields demonstrates a ground state with $S > 0$. At intermediate fields a clear torque step is seen indicative of a level-crossing at a field $B^* \approx 7.5 \,$T \cite{CsFe8,XFe6}. From $B^*$ the excitation energy of the excited state involved in the level-crossing is estimated to $\Delta_1 \approx 10 \,$K. The anisotropy of the ground state is too weak to yield such excitation energies (see below). The level-crossing thus occurs between different spin levels with $| \Delta S | \ge 1$. However, the most astonishing feature of the data is the sign change of the torque signal at the level-crossing, i.e. a change of the magnetic anisotropy from easy-axis to hard-axis type. This is immediately evident from Fig. 3(a) as the magnitude of the torque is related to the magnetic anisotropy \cite{CsFe8,XFe6}. These observations actually represent the main result of this work. 

\begin{figure}
\includegraphics{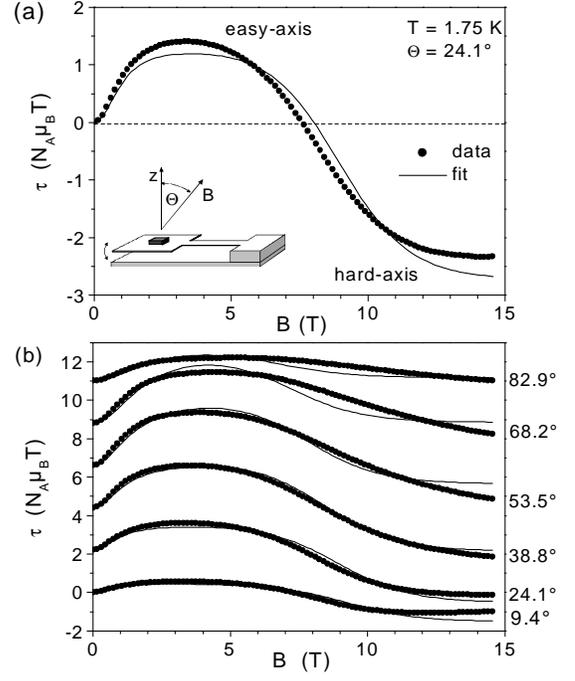}
\caption{
Field dependencies of the torque for a Mn-[3$\times$3] single crystal at various angles. The solid curves represent a fit with the model described in the text ($D_0=-0.5\,$K, $B_0=0.005\,$K, $\Delta_1=10.2\,$K, $D_1=0.1\,$K, $B_1=0.001\,$K). The inset shows a sketch of the silicon-cantilever torquemeter with a sample mounted. It also defines the $z$-axis and the angle $\Theta$.
}
\end{figure}

%

Assuming an idealized [3$\times$3] grid structure, the appropriate spin Hamiltonian for the 
theoretical interpretation

\begin{eqnarray}
\label{eq1}
H = 
- J_R ( \sum_{i=1,...,7} {\bf S}_i \cdot {\bf S}_{i+1} + {\bf S}_8 \cdot {\bf S}_1 ) \cr 
- J_C ( {\bf S}_2 + {\bf S}_4 +{\bf S}_6 + {\bf S}_8 ) \cdot {\bf S}_9 
+ \sum_{i<j} {\bf S}_i \cdot {\bf D}_{ij}^{dip} \cdot {\bf S}_j \cr
+ \sum_i {\bf S}_i \cdot {\bf D}_i^{lig} \cdot {\bf S}_i
+ \mu_B g \sum_i {\bf S}_i \cdot {\bf B}
\end{eqnarray}

consists of the isotropic nearest-neighbor exchange terms, the dipole-dipole interaction terms, the second order ligand-field terms and the Zeeman terms. $J_R$ characterizes the couplings of the eight outer Mn ions, $J_C$ those involving the central Mn ion. The first two terms on the rhs of eq. (1) will be abbreviated as $H_R$ and $H_C$, respectively. For the Mn-[3$\times$3] grid, $S_i = 5/2$.

The analysis of the data in the sense of determining the parameters of eq. (1) is hampered by the 
huge dimension of the Hilbert space of 10$\,$077$\,$696. Even if only the isotropic exchange 
terms are considered and all symmetries (i.e. spin rotational and $D_4$ spin permutational 
symmetry \cite{Symmetrie}) are exploited, the largest dimension is still 22$\,$210. This exceeded 
our computer capabilities. Thus, in the following the data will be discussed at various levels of 
approximation. They are too crude to allow for a quantitative analysis, but suggest a mechanism for 
the sign change of the magnetic anisotropy.

Hamiltonian (1) bears a close relation to that of a ring of eight $S_i = 5/2$ spin centers. The rhs of eq. (1) may be split into terms involving only the "ring" of eight outer Mn ions, terms related to the central Mn ion, and terms representing an interaction between these two sets of ions. Fortunately, the problem of the octanuclear ring has been completely solved in the strong exchange limit recently \cite{CsFe8}, and we will profit greatly from these results.

The decrease of $\chi T$ towards low temperatures seen in Fig. 2(a) requires at least one of the 
coupling constants $J_R$ and $J_C$ to be antiferromagnetic; the $S = 5/2$ ground state tells that  $J_R$ must be antiferromagnetic. The solid line in Fig. 2(a) represents a fit using the 
theoretical $\chi T$ curve for the octanuclear ring with the addition of 7.84$\, N_{A} \, \mu_B \,$T$^{-1} \,$K to account for the additional central Mn ion. It corresponds to $J_C = 0$. The agreement with data is surprisingly good with $J_R$ = -5.5$\,$K. However, the good agreement does not mean that 
$J_C$ is really zero. Numerical studies for [3$\times$3] grids with spin-2 centers, for which 
$\chi T$ could be calculated exactly, showed that for $J_C/|J_R| \in [-1,0.5]$ the theoretical $\chi T$ curves can be matched almost perfectly to the one with $J_C = 0$, if $J_R$ is renormalized 
appropriately [Fig. 2(a)]. Deviations are visible for temperatures $T \approx |J_R|$, but are too small to be detectable by our experiment. Actually, if $J_C = 0$ one would calculate a value of 
3$\,$K for the gap $\Delta_1$ which is considerably smaller than the observed value of 10$\,$K. This is attributed to a significant renormalization of $J_R$ due to a non-zero $J_C$.

Since the anisotropic terms are small for high-spin Mn(II) ions, Hamiltonian (1) may be analyzed in 
the strong exchange limit. The anisotropic corrections are then treated in first-order perturbation 
theory yielding the effective spin Hamiltonian

\begin{equation}
\label{eq2}
H_n = \Delta_n {\bf 1} + {\bf S} \cdot {\bf D}_n \cdot {\bf S} + \mu_B g {\bf S} \cdot {\bf B}
\end{equation}

for each eigenstate of $H_R+H_C$ \cite{Bencini90}. The quantities $\Delta_n$ and ${\bf D}_n$ are related to the microscopic parameters of eq. (1) via $\Delta_n = - \sum_{i<j} a^n_{ij} J_{ij}$ and 

\begin{equation}
\label{eq3}
{\bf D}_n = \sum_{i<j} b^n_{ij} {\bf D}^{dip}_{ij} + \sum_{i} c^n_i {\bf D}^{lig}_i.
\end{equation}

The projection coefficients $a$, $b$, and $c$ may be calculated from the eigenvectors - if known. If not, as in the present case, eq. (2) provides at least a starting point for a phenomenological description of the data.

The magnetization curve could be reproduced including three levels with $S = 5/2, 7/2, 9/2$. The resulting curve and values for $D_0$, $\Delta_1$, $D_1$, $\Delta_2$, and $D_2$ are displayed in Fig. 2(b). The meaning of the parameters is clarified in Fig. 4. With the same approach we fitted the torque data. For the fit, the data of field-sweep measurements for 15 angles were used simultaneously (not all shown in Fig. 3). The same model as used for the magnetization curve reproduced the data only roughly. This demonstrates that higher-order terms are non-negligible. We thus included a forth-order ligand-field term, $B_n {\bf O}_{40}({\bf S})$ \cite{Bencini90}, in eq. (2) which is expected to be the most important higher-order term. The solid curves in Fig. 3 represent a result where only two $S = 5/2, 7/2$ levels were considered. The discrepancies visible at the highest field indicate the beginning influence of a $S = 9/2$ level. Its inclusion led to better fits, but with this many parameters the fitting routine converged badly. Accordingly, we did not try to expand eq. (2) with even more terms.

Evidently, the sign change of the torque can be related to opposite signs of $D_0$ and $D_1$, as expected. But in view of eq. (3) this cannot be simply a result of different signs of the microscopic ligand-field terms ${\bf D}^{lig}_i$. It must be related to the projection coefficients $b^n_{ij}$ and in particular to $c^n_i$. To understand the sign change it is thus necessary to get at least an idea of their values. 

The good fit of the magnetic susceptibility by the $J_C$ = 0 curve suggests a perturbational treatment in the limit $|J_C/J_R| \ll 1$. $H_R$ is regarded as the dominant term and remaining terms are treated in first-order perturbation theory. This effectively replaces the [3$\times$3] grid by a "dimer" composed of the spin of the outer ring of Mn ions and the spin of the central Mn ion coupled via $J_C$ (and dipole-dipole terms).

The uncoupled wave functions are $| \alpha S_R M_R \rangle \cdot |S_9 M_9 \rangle$. $\alpha$ abbreviates quantum numbers needed for an unambiguous classification of the states of the ring. It will be omitted in the following. Switching on $J_C$ leads to an admixture of these wave functions and the zero-order wave functions are obtained by coupling them to $|S_R S_9 S M \rangle$ according to ${\bf S} = {\bf S}_R + {\bf S}_9$ \cite{Bencini90}. The resulting spectrum is schematically shown in Fig. 4. The $S_R = 0$ ground state of the ring couples with the $S_9=5/2$ spin of the central ion to give the $S = 5/2$ ground state of the [3$\times$3] grid. The first excited state of the ring with $S_R = 1$ couples to $S = 3/2, 5/2, 7/2$, the latter being the state relevant for the level-crossing visible in the torque data. 

\begin{figure}
\includegraphics{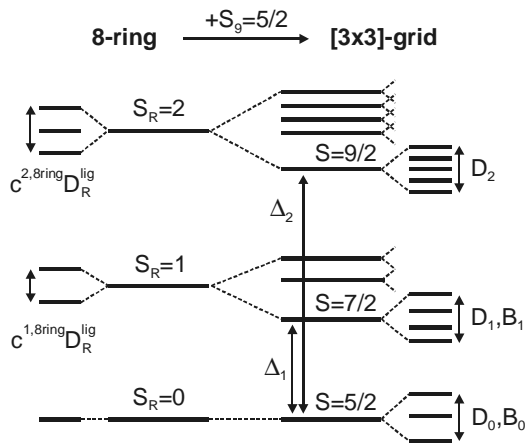}
\caption{
Sketch of the energy spectrum for the [3$\times$3]-grid as it appears in the "dimer" approximation discussed in the text. The states of the ring of Mn ions (shown on the left) are coupled with the state of the central Mn ion to yield the states of [3$\times$3]-grid (shown on the right). The meaning of some of the parameters discussed in the text is sketched.
}
\end{figure}

The quantity $c^n_i$ is given by the projection of the single ion state $|S_i M_i \rangle$ onto the grid-state $| S_R S_9 S M \rangle$. For the central ion this yields exactly the projection coefficient $c^{n,dimer}_2$ of a dimer with spins $S_R$ and $S_9$ \cite{Bencini90}. For an ion of the ring, the projection may be done in two steps. One first projects $|S_i M_i \rangle$ onto the ring-state $|S_R M_R \rangle$, and then $|S_R M_R \rangle$ onto $|S_R S_9 S M \rangle$. The first part is equal to calculating the projection coefficients for the ring. This, fortunately, has been done already in Ref. \onlinecite{CsFe8}; the corresponding coefficients will be denoted as $c^{n,ring}$. The second part leads to $c^{n,dimer}_1$, also a dimer coefficient. Thus, $c^n_9 = c^{n,dimer}_2$ and $c^n_R = c^{n,dimer}_1 c^{n,8ring}$; the subscript "$R$" refers to one of the eight ions on the ring. 

Table I shows that, while the dimer coefficients are positive, the ring coefficient is zero for the $S = 5/2$ ground state and negative (!) for the next higher states. Quantitatively, it holds that $D_0 = D^{lig}_9$ and $D_1 = -0.78 D^{lig}_R + 0.48 D^{lig}_9$.  $D^{lig}_R$ denotes the average contribution of an ion of the ring \cite{CsFe8}. The minus sign in the equation for $D_1$ provides an explanation for the opposite signs of $D_0$ and $D_1$. 

A similar analysis was put through for the dipole-dipole contribution $D^{dip}_n = \sum_{i<j} b^n_{ij} D^{dip}_{ij}$. It is calculated that $D^{dip}_n = 0.05\,$K, which should be compared to $D^{lig}_9 \approx -0.5\,$K. It thus does not invalidate the argument.

\begin{table}
\caption{Values of projection coefficients discussed in the text for the lowest levels with 
$S=5/2,7/2,9/2$.}
\label{table1} 
\begin{ruledtabular}
\begin{tabular}{ccccccc}
8-ring & $c^{n,ring}$ && [3$\times$3]-grid & $c^{n,dimer}_1$ & $8c^n_R$ & 
$c^{n,dimer}_2$ \\
\tableline
$S_R=0$ &  0          & & $S=5/2$ & 0      &   0          &   1     \\
$S_R=1$ & -16.348 & & $S=7/2$ & 1/21 & -0.7785 & 10/21 \\
$S_R=2$ & -3.8080 & & $S=9/2$ & 3/18 & -0.6347 & 5/18   \\
\end{tabular} 
\end{ruledtabular}
\end{table} 

In view of the involved approximations, the analysis presented here cannot claim to be quantitative. But it provides a possible mechanism for the sign change of the magnetic anisotropy: The anisotropy of the $S = 5/2$ ground state is dominated by that of the central ion, while the anisotropy of the first excited $S = 7/2$ level is additionally controlled by that of the ring of outer ions. The sign change arises since the effective anisotropy of the ring is of opposite sign to that of the involved microscopic single-ion anisotropies.

%

In conclusion, we found the supramolecular Mn(II)-[3$\times$3] grid [Mn$_9$(2POAP-2H)$_2$]$^{6+}$ to be of interest due to the following points: (i) It is the first example of a molecular nanomagnet with a ground state with $S > 0$ for which a level-crossing involving states with different $S$ could be observed. (ii) The sign of the magnetic anisotropy changes abruptly at the level-crossing, i.e. can be controlled by the magnetic field. (iii) The Mn-[3$\times$3] grid was regarded as a dimer composed of an octanuclear ring and a single Mn ion. This suggests that although a [3$\times$3] grid topology is clearly "two-dimensional" it preserves characteristics of one-dimensionality.

%

%

%
\end{document}